\documentclass[%
superscriptaddress,
preprint,
 amsmath,amssymb,
 aps,
prb,
]{revtex4-2}

\usepackage{graphicx}
\usepackage{dcolumn}
\usepackage{bm}
\usepackage{subcaption}

\begin{document}


\title{Pressure tuning of putative quantum criticality on YbV$_6$Sn$_6$}

\author{P. C. Sabino}
\affiliation{CCNH, Universidade Federal do ABC (UFABC), Santo André, SP, 09210-580, Brazil}
\author{L. Mendonça-Ferreira}%
\affiliation{CCNH, Universidade Federal do ABC (UFABC), Santo André, SP, 09210-580, Brazil}
\author{J. G. Dias}
\affiliation{CCNH, Universidade Federal do ABC (UFABC), Santo André, SP, 09210-580, Brazil}
\author{G. G. Vasques}
\affiliation{CCNH, Universidade Federal do ABC (UFABC), Santo André, SP, 09210-580, Brazil}
\author{M.  Dutra}
\affiliation{CCNH, Universidade Federal do ABC (UFABC), Santo André, SP, 09210-580, Brazil}
\author{H. Pizzi}
\affiliation{Instituto de Física Gleb Wataghin, UNICAMP, 13083-859, Campinas, SP, Brazil}%
\author{P. G. Pagliuso}
\affiliation{Instituto de Física Gleb Wataghin, UNICAMP, 13083-859, Campinas, SP, Brazil}%
\author{M. A. Avila}%
\email{avila@ufabc.edu.br}
\affiliation{CCNH, Universidade Federal do ABC (UFABC), Santo André, SP, 09210-580, Brazil}

%
%


\begin{abstract}

YbV$_6$Sn$_6$ is a recently discovered heavy-fermion compound that orders at T$_N$ $\approx$ 0.4~K and exhibits a magnetic field-tuned quantum critical point at H $\approx$ 10 kOe.
In this work, we have grown YbV$_6$Sn$_6$ single crystals by the self-flux method, to investigate their physical properties at ambient pressure and their electrical transport properties under hydrostatic pressure.
At higher temperatures, we observed a decrease in the Kondo temperature, accompanied by the appearance of a local minimum followed by a local maximum, associated with the onset of the coherent Kondo regime.
Power law fitting at low temperatures indicated a recovery of the Fermi-liquid regime for pressures below 1 GPa.
Above 1 GPa, a reentrance of non-Fermi-liquid behavior is suggested by a decrease in the exponent $n$, accompanied by a substantial increase in the parameter A, indicating the approach of a new quantum criticality tuned by hydrostatic pressure.
The broad range of interactions present in YbV$_6$Sn$_6$, including RKKY, crystalline electric field (CEF), and Kondo lattice effects, appears to lead to a complex phase diagram.
we present a putative phase diagram featuring double quantum criticality tuned by both magnetic field and hydrostatic pressure.

\end{abstract}

\maketitle


\section{Introduction}

Heavy-fermion (HF) compounds continue to present significant challenges and opportunities in condensed matter physics, owing to the strong correlation between conduction electrons and localized $f$-electron moments \cite{coleman2007heavy}.
Their characteristically small energy scales allow for convenient tuning of ground states, making them ideal systems for investigating quantum critical points (QCPs) induced by pressure, magnetic field, and chemical substitution \cite{doi:10.1126/science.1191195}.
Near QCPs, dominant quantum fluctuations give rise to highly collective excitations and exotic quantum phases, such as unconventional superconductivity \cite{BATLOGG1987441}.

The classical Doniach phase diagram provides a foundational framework \cite{doniach1977kondo}, describing the competition between the Kondo effect and the Ruderman-Kittel-Kasuya-Yosida (RKKY) interaction in determining the ground state of HF compounds.
While this scenario typically predicts a single QCP between an antiferromagnetic (AFM) ordering and a heavy Fermi-liquid (FL) state, recent experimental studies have revealed its insufficiency for describing the diverse quantum critical behaviors in many HF materials, particularly where frustration effects play a significant role \cite{Yang_2023}.
This has motivated the development of more complex global phase diagrams that incorporate the interplay between the Kondo effect and magnetic frustration.

Ytterbium (Yb)-based heavy-fermion compounds are of particular interest.
While often less common than their cerium (Ce)-based counterparts, partly due to synthesis challenges associated with Yb's high vapor pressure \cite{fisk1992existence}, they can offer unique insights into complex quantum phenomena.
Notably, unconventional superconductivity and non-Fermi-liquid (NFL) behavior have been observed in Yb-based systems such as YbAlB$_4$ \cite{nakatsuji2008superconductivity} and YbRh$_2$Si$_2$ \cite{schuberth2016emergence}.
Furthermore, magnetic fields can induce various quantum critical states and multiple phase transitions in these materials, advancing the understanding of NFL states, Fermi surface reconstruction, and global phase diagrams in HF systems.

The \textit{RT}$_6$\textit{X}$_6$ family of intermetallic compounds (\textit{R} = rare earth, \textit{T} = transition metal, \textit{X} = Ge, Sn) provides an excellent structural platform for exploring these phenomena \cite{xu2023quantum}.
Hereafter referred to as the \textit{R}166 structure, it is characterized by a Kagome lattice of transition metal atoms and a triangular lattice of rare-earth atoms.
The great interest in Kagome materials stems from their potential for hosting emergent strongly correlated and topological physics \cite{wang2023quantum, hu2022tunable, peng2021realizing, pokharel2021electronic, arachchige2022charge}.
A crucial aspect contributing to such is the hybridization and interaction between the $d$-electrons of the transition metal and the $f$-electrons of the rare-earth element, which strongly influences the competition between Kondo screening and magnetic order.
While the broad range of compounds in \textit{R}166 family has been extensively studied, particularly for their structural and Kagome-related topological features, the strong $4f$-electron correlation in Yb-based members remains much less explored.

In this work, we present a comprehensive study on single crystals of YbV$_6$Sn$_6$, a new Yb-based R166 compound featuring a triangular Kondo lattice \cite{YbV6Sn6}.
Our electrical resistivity measurements under hydrostatic pressures suggest the possibility of a second quantum critical point in YbV$_6$Sn$_6$ tuned by pressure.
These findings, together with magnetic susceptibility and heat capacity data which confirm the HF ground state and aid in understanding CEF effects, suggest the presence of an enriched phase diagram that may offer new insights into the multi-parameter tuning of quantum criticality, as well as the interplay between Kondo physics, magnetism, and potentially other emergent phases in Yb-based heavy-fermion systems.

\section{Experimental Methods}

High-quality single crystals of the YbV$_6$Sn$_6$ were grown by the self-flux method, similar to previously reported procedure \cite{YbV6Sn6}.
The starting materials were Yb (Ames), V (Matek, 99.99\%) and Sn (99.999\%), mixed in the molar ratio of 1:6:20, then inserted and sealed in a quartz ampoule under vacuum.
The ampoule was heated to 500~$^\circ$C, dwelled for 1~h, then heated up to 1150~$^\circ$C in 2~h and dwelled for 24~h to ensure a homogeneous melt of reagents.
Subsequently, they were cooled at a constant rate of 1~$^\circ$C/h to 780~$^\circ$C and finally centrifuged to remove excess of Sn flux.
HCl was later used to remove the remaining flux from crystals, which have typical dimensions of $0.7 \times 0.5 \times 0.1$~mm$^3$.
The crystal structure was confirmed through powder X-ray diffraction (PXRD) on crushed crystals using a STOE STADI-P diffractometer with Cu-K$\alpha$ radiation.
Elemental analysis was performed by energy dispersive x-ray spectroscopy (EDX) on a JEOL JSM-6010LA, coupled to a scanning electron microscope (SEM).
Magnetic measurements were made on a Quantum Design MPMS3 SQUID-VSM.
Heat capacity and four-probe resistivity were carried out on a Quantum Design PPMS EverCool II system.
DC electrical resistivity in the temperature range $2\leq T \leq 300$~K and $0 \leq P \leq 2.8$~GPa was measured in a hydrostatic pressure cell (HPC) with silicon oil as pressure-transmitting medium.
The pressure was determined from the pressure dependence of the superconducting transition temperature of Pb.

\section{Results and Discussion}

\subsection{Crystal Structure}

\begin{figure}[ht!]
    \centering
    \includegraphics[width=0.9\columnwidth]{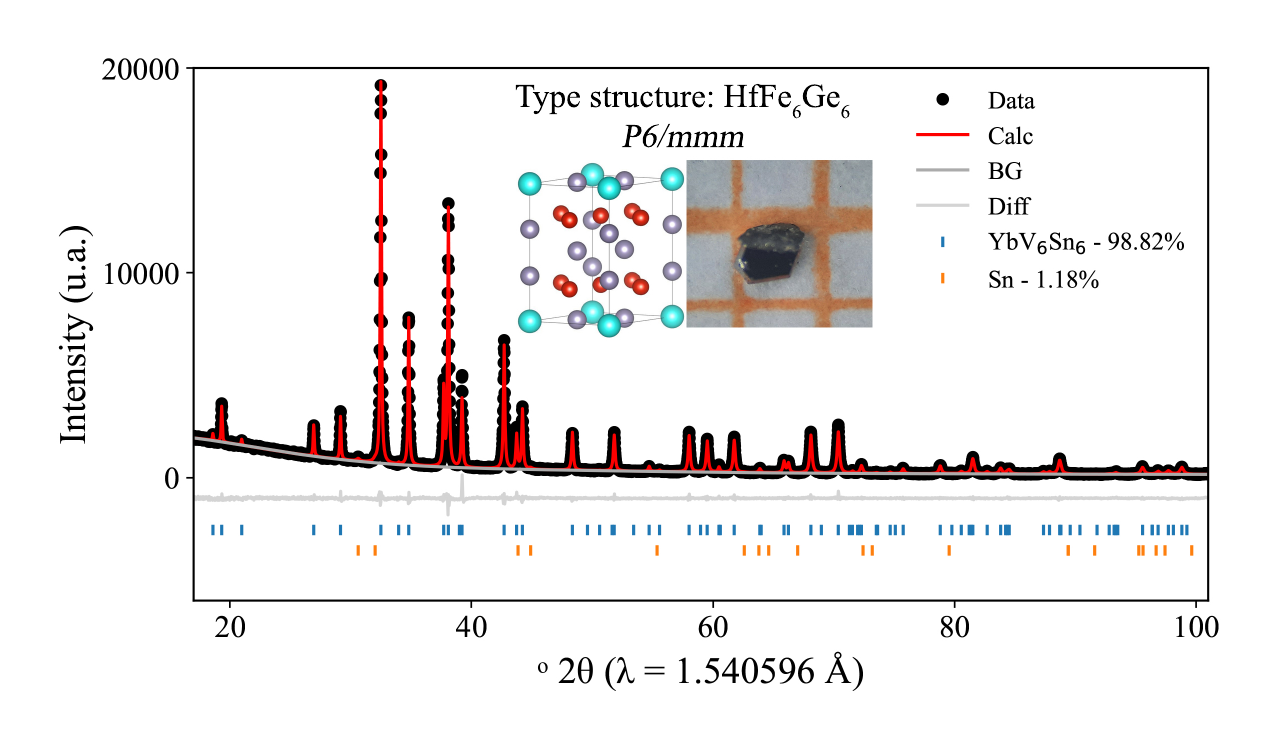}
    \caption{Room temperature powder X-ray diffraction (PXRD) pattern of YbV$_6$Sn$_6$ single crystals, measured using Cu-K$\alpha$ radiation. The black circles represent the experimental data, the red line is the Rietveld refinement, the dark grey line is the background, and the light grey line at the bottom shows the difference curve. The vertical ticks indicate the Bragg peak positions for YbV$_6$Sn$_6$ (blue) and a residual amount of Sn flux (orange, $\approx$1.18\% ). The inset shows a representation of the YbV$_6$Sn$_6$ crystal structure (HfFe$_6$Ge$_6$-type, space group P6/mmm) and a typical as-grown single crystal.}
    \label{fig:GMQ_926_Diffraction}    
\end{figure}

Figure \ref{fig:GMQ_926_Diffraction} shows the X-ray diffraction pattern for YbV$_6$Sn$_6$ together with the Rietveld refinement, which confirms that our samples crystallize in the \textit{P6/mmm} space group the HfFe$_6$Ge$_6$-type structure, corresponding to the structurally ordered phase of the \textit{R}166 family.
Powder diffraction analysis showed no evidence of disordered variants, such as the Yb$_{0.5}$Co$_3$Ge$_3$- or SmMn$_6$Sn$_6$-type structures.

The lattice parameters obtained from Rietveld refinement are listed in Table \ref{tab:PXRD_par} and exhibit a deviation of less than 0.1\% from literature \cite{YbV6Sn6}.
Additionally, the presence of minor peaks associated with $\beta$-Sn as a secondary phase ($\approx$ 1.18\%) was observed.
This phase originates from residual flux attached on the crystal surfaces and was removed prior to further measurements using a pure HCl solution.

\begin{table}[ht!]
\centering
\caption{Lattice parameters, unit cell volumes, atomic coordinates and isotropic displacement parameters of YbV$_6$Sn$_6$ at 300 K.}
\label{tab:PXRD_par}
\begin{tabular}{l|c|ccc|c}
\multicolumn{1}{c|}{Lattice Parameters} & Atom site & \multicolumn{3}{c|}{Position}                                             & B$_\textnormal{ISO}$ \\ \hline
\multicolumn{1}{c|}{P6/mmm (\#191)}      &           & \multicolumn{1}{c|}{x}       & \multicolumn{1}{c|}{y}       & z           &                      \\ \hline
a = 5.50364(8) \AA                      & Yb (1a)   & \multicolumn{1}{c|}{0}       & \multicolumn{1}{c|}{0}       & 0           & 2.56(4)              \\ \hline
c = 9.17766(15) \AA                     & V (6i)    & \multicolumn{1}{c|}{\ \ \ \ \ 1/2\ \ \ \ \ }     & \multicolumn{1}{c|}{0}       & 0.24862(19) & 1.82(4)              \\ \hline
V = 240.748 \AA$^3$                     & Sn1 (2c)  & \multicolumn{1}{c|}{\ \ \ \ \ 1/3\ \ \ \ \ } & \multicolumn{1}{c|}{\ \ \ \ \ 2/3\ \ \ \ \ } & 1/2         & 1.96(4)              \\ \hline
                                        & Sn2 (2d)  & \multicolumn{1}{c|}{1/3} & \multicolumn{1}{c|}{2/3} & 0           & 1.79(3)              \\ \hline
GOF = 1.193                             & Sn3 (2e)  & \multicolumn{1}{c|}{0}       & \multicolumn{1}{c|}{0}       & 0.32991(13) & 1.99(3)             
\end{tabular}
\end{table}

It is noteworthy in Table~\ref{tab:PXRD_par} that the isotropic displacement parameter (B$_\textnormal{ISO}$) for Yb (1a) has an unusually elevated value compared to the others.
This finding contrasts with the expected behavior for the insertion of rare-earth ions in a CoSn-type structure, where the rare-earth ion should typically shift the Sn3 (2e) atoms, becoming entrapped in a cage formed by Sn and V atoms \cite{Venturini+2006+511+520}.
Here, the observed B$_\textnormal{ISO}$ value suggests a certain degree of mobility for the Yb ion.

Alternatively, the Sn3 (2e) atoms might also be loosely bound within the structure.
In this scenario, due to the influence of the Yb atom, an anisotropic displacement predominantly in the \textit{c}-direction could be present for Sn3 (2e).
As the current analysis is isotropic, the B$_\textnormal{ISO}$ value might mask this potential looseness.
To further elucidate this issue, temperature-dependent measurements incorporating anisotropic displacement analysis are being scheduled.

\subsection{Magnetization}

\begin{figure}[ht!]
    \centering
    \includegraphics[width=0.9\columnwidth]{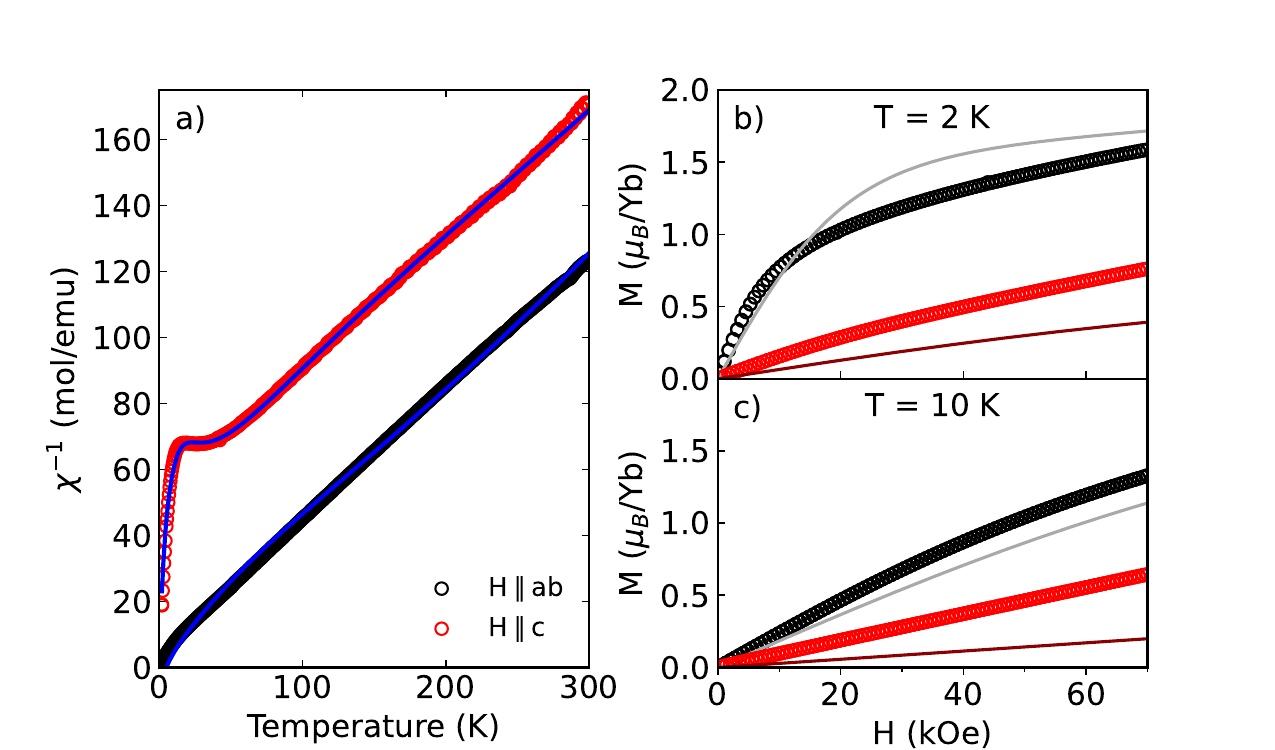}
    \caption{ (a) Inverse magnetic susceptibility ($\chi^{-1}$) as a function of temperature for YbV$_6$Sn$_6$ measured with an applied magnetic field of \textit{H}=10 kOe, for $H\parallel c-axis$ (red open circles) and $H\parallel ab-plane$ (black open circles). The solid lines represent fits considering crystalline electric field (CEF) effects. (b) Isothermal magnetization (\textit{M}) as a function of magnetic field (\textit{H}) at T=2 K and (c) T=10 K, for $H\parallel c-axis$ (red circles) and $H\parallel ab-plane$ (black circles).}
    \label{fig:GMQ_926_Mag}    
\end{figure}

Figure \ref{fig:GMQ_926_Mag}a displays the inverse magnetic susceptibility for applied fields of 10 kOe along the crystallographic \textit{c}-direction and the \textit{ab}-plane.
Above 200~K the response follows Curie-Weiss behavior, giving the magnetic moments and Weiss paramagnetic temperatures shown in Table \ref{tab:CW-fit}.
The negative $\theta_{CW}$ indicates predominance of antiferromagnetic (AFM) interactions in both directions, with an effective magnetic moment expected of Yb$^{3+}$ ($\mu_{eff}$ = 4.54 $\mu_B$) \cite{mugiraneza2022tutorial}.
Figs.~\ref{fig:GMQ_926_Mag}b and \ref{fig:GMQ_926_Mag}c show the isothermal magnetization responses at 2~K and 10~K, respectively.

\begin{table}[ht!]
\centering
\caption{Parameters obtained from the Curie-Weiss fitting of magnetic susceptibility data in the high-temperature regime (200-300~K) for YbV$_6$Sn$_6$, showing values for H$\parallel c$, H$\parallel ab$, and the polycrystalline average. The parameters include the Curie-Weiss temperature ($\theta_{CW}$), effective magnetic moment ($\mu_{eff}$), and temperature-independent susceptibility ($\chi_0$).}
\label{tab:CW-fit}
\begin{tabular}{l|c|c|c}
\multicolumn{1}{c|}{}               & $ H \parallel c $ & $H \parallel ab $ & Average  \\ \hline
$\theta_{CW}$ (K)                   & -136.5(9)         & -19.4(3)          & -39.4(2) \\ \hline
$\mu_{\textnormal{eff}}$  ($\mu_B$) & 4.543(6)          & 4.537(3)          & 4.533(2) \\ \hline
$\chi_0$ (emu/mol)                  & 0.03749(1)        & -0.001318(7)      & --      
\end{tabular}
\end{table}

At high temperatures, there is significant magnetic anisotropy due to the CEF associated with Yb$^{3+}$ ions.
A local maximum around 31.2~K, followed by a local minimum near 11.5~K for $H~\parallel~c$, is also attributed to CEF effects and has been observed in other members of the RV$_6$Sn$_6$ family \cite{RV6Sn6_magnetic}, depending on the hard magnetization axis, except for R = Gd, Dy, and Ho.  
For $H \parallel ab$, a significantly stronger magnetic response is observed, consistent with the tendency for an easy magnetization plane in rare-earth elements with smaller atomic radii (Er and Tm) \cite{RV6Sn6_magnetic, RV6Sn6_electronic}.

In order determine the CEF parameters, magnetic susceptibility ($\chi_c$ and $\chi_{ab}$) and isothermal magnetization ($M_c$ and $M_{ab}$ under various temperatures) was computed using the Hamiltoninan \cite{ARNOLD2014156}:
\begin{equation}
    H = H_{CEF} + g_{J}\mu_{B}\textbf{J}\cdot\textbf{B}_{\textnormal{EXT}} + 2(g_{J}-1)\mu_{B}\textbf{J}\cdot\textbf{B}_{\textnormal{MOL}}\ ,
\end{equation}
where $g_J$ is the Lande g-factor, $\mu_B$ is the Bohr magneton, $\textbf{B}$ is the magnetic field ( external and molecular) and $\textbf{J}$ is the total angular momentum operator vector.
The first term is the CEF Hamiltonian and it is defined for a $D_{6h}$ point symmetry as:
\begin{equation}
    H_{CEF} = B^0_2O^0_2 + B^0_4O^0_4 + B^0_6O^0_6 + B^6_6O^6_6\ ,
\end{equation}
where $O^m_n$ are the Stevens operators and $B^m_n$ characterize the crystal-field parameters.

Similarly to Curie-Weiss susceptibility, spin-spin interactions were considered using a parameter $\lambda$.  
The results are presented in Table \ref{tab:VanVleck_fit}.
In contrast to the Curie-Weiss (CW) fitting, the $\lambda_{c,ab}$ values obtained from the simultaneous fits suggest a predominance of ferromagnetic (FM) interactions among the Yb ions.
The reason for this discrepancy is not yet clear and will require a more detailed analysis of the spin-spin interactions, especially considering that the previously reported magnetic ordering at 0.4~K \cite{YbV6Sn6} is of the antiferromagnetic (AFM) type.

\begin{table}[ht!]
\centering
\caption{CEF parameters ($B^m_n$) and the corresponding energy level splittings ($\Delta_i$) for Yb$^{3+}$ ions in YbV$_6$Sn$_6$ plus the molecular field parameters ($\lambda_{ab}$ and $\lambda_{c}$).}
\label{tab:VanVleck_fit}
\begin{tabular}{lc|l|lc}
\multicolumn{2}{c|}{CEF parameters}                                      &  & \multicolumn{2}{c}{Energy splitting}         \\ \cline{1-2} \cline{4-5} 
\multicolumn{1}{l|}{$B^0_2$ (K)}              & 6.93 $\times 10^{-1}$  &  & \multicolumn{1}{l|}{$\Delta_1$ (K)} & 7  \\ \cline{1-2} \cline{4-5} 
\multicolumn{1}{l|}{$B^0_4$ (K)}              & 9.42 $\times 10^{-3}$  &  & \multicolumn{1}{l|}{$\Delta_2$ (K)} & 23 \\ \cline{1-2} \cline{4-5} 
\multicolumn{1}{l|}{$B^0_6$ (K)}              & -8.00 $\times 10^{-4}$ &  & \multicolumn{1}{l|}{$\Delta_3$ (K)} & 65 \\ \cline{1-2} \cline{4-5} 
\multicolumn{1}{l|}{$B^6_6$ (K)}              & -3.27 $\times 10^{-2}$ &  & \multicolumn{1}{l|}{}               &        \\ \cline{1-2} \cline{4-5} 
\multicolumn{1}{l|}{$\lambda_{ab}$ (mol/emu)} & 5.23                   &  & \multicolumn{1}{l|}{}               &        \\ \cline{1-2} \cline{4-5} 
\multicolumn{1}{l|}{$\lambda_{c}$ (mol/emu)}  & 2.51                   &  & \multicolumn{1}{l|}{}               &       
\end{tabular}
\end{table}


\subsection{Heat Capacity}

\begin{figure}[ht!]
    \centering
    \includegraphics[width=0.9\textwidth]{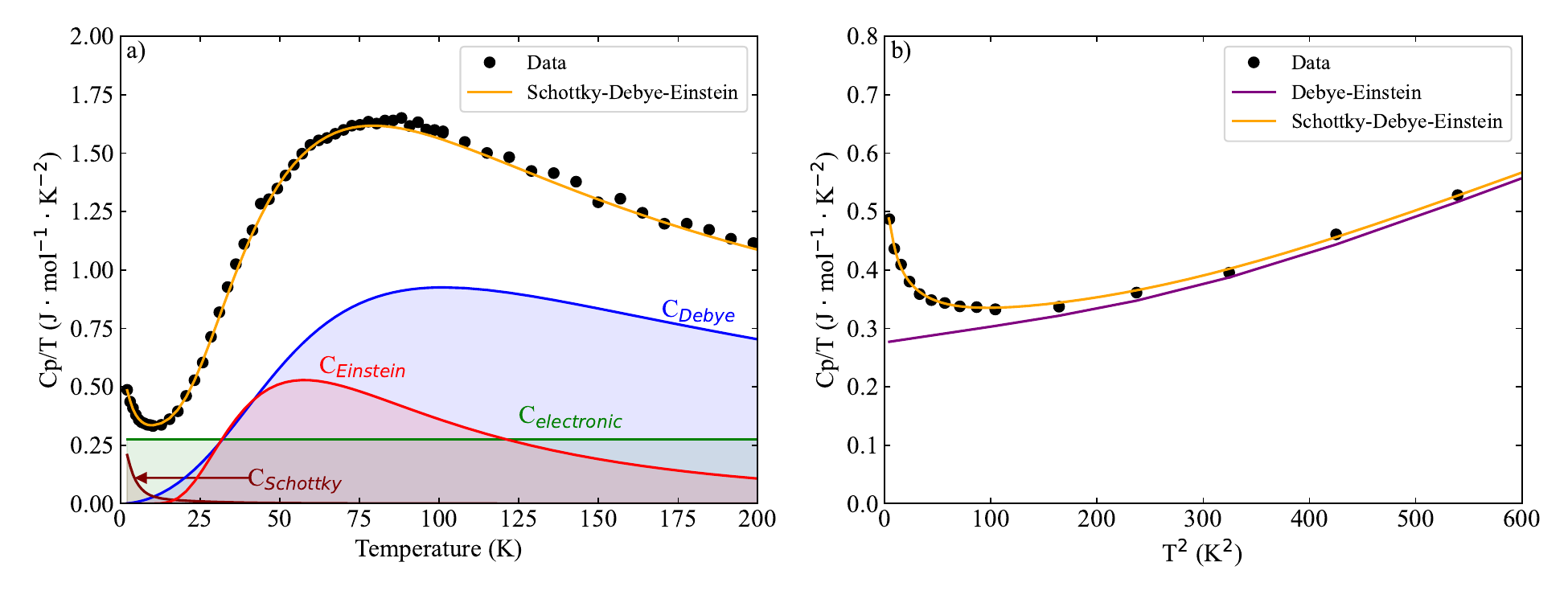}
    \caption{Temperature dependence of the specific heat divided by temperature (Cp/T) for YbV$_6$Sn$_6$. Black circles represent the experimental data. The orange line is the total fit considering electronic (C$_{el}$), Schottky (C$_{Schottky}$), Debye (C$_{Debye}$), and Einstein (C$_{Einstein}$) contributions. The individual contributions are shown as shaded areas: electronic (green), Schottky (maroon), Einstein (red), and Debye (blue). (b) Low temperatures dependence with the resulting fit for electronic and phonon contributions (purple) and considerating the Schottky contribution (orange).}
    \label{fig:GMQ_926_HC}    
\end{figure}

Figure \ref{fig:GMQ_926_HC} presents the specific heat data in (a) zero field in the temperature range from 2~K to 200~K and (b) from 2~K to 25~K.
A noticeable increase in specific heat is observed with decreasing temperature below 10~K.
Given that this temperature is too high for a nuclear Schottky contribution, this increase is attributed to the Schottky anomaly generated by CEF splitting.

A fit was performed between 10~K and 200~K considering both the electronic and phononic contributions to the specific heat in the form: $C = C_{el} + C_{ph}$.
In this fit, the Debye model alone proved insufficient for a good fit of the experimental data. Considering the high value of the isotropic displacement parameter (B$_\textnormal{ISO}$) for Yb atoms obtained from the Rietveld refinement, an Einstein contribution term was added to the phononic part, suggesting possible \textit{rattling} of these atoms.
With a considerably improved fit, the parameters presented in Table \ref{tab:GMQ_926_HC} were obtained.

The resulting Sommerfeld coefficient ($\gamma$ = 278(1) mJ mol$^{-1}$ K$^{-2}$) differs from the value of 411 mJ mol$^{-1}$ K$^{-2}$ reported by Guo \textit{et al.} \cite{YbV6Sn6}, but it remains three orders of magnitude larger than that of conventional metals like Cu, confirming heavy fermion behavior.

\begin{table}[ht!]
\centering
\caption{Parameters obtained from the fit of the specific heat data, including the Debye term coefficient (D), Debye temperature ($\theta_D$), Einstein term coefficient (E), Einstein temperature ($\theta_E$), and Sommerfeld coefficient ($\gamma$).}
\label{tab:GMQ_926_HC}
\begin{tabular}{l|c}
\multicolumn{1}{c|}{ Parameter} & Results  \\ \cline{1-2} 
$D$                                                                           & 0.508(4)                          \\
$\theta_D$ (K)                                                                & 362(2)                            \\
$E$                                                                           & 0.0284(3)                         \\
$\theta_E$ (K)                                                                & 173(1)                          \\
$\gamma$ (mJ $\cdot$ mol$^{-1}$ $\cdot$ K$^{-2}$)                              & 278(1)                          \\ \cline{1-2}
\end{tabular}
\end{table}

The Schottky fit was performed using a multi-level model \cite{souza2016specific}, given the three low-lying Kramers doublets from the CEF splitting of the Yb$^{3+}$ $4f$ levels.
While we successfully obtained splitting energies of $\Delta_1=6(1)$~K and $\Delta_2 = 18(2)$~K by adjusting the parameters, the fitting for $\Delta_3$ consistently diverged.
This suggests that phononic contributions might be masking the Schottky anomaly near 64~K.
Despite this, the excellent overall fit confirms that the system maintains at least two Yb$^{3+}$ CEF degeneracies.
Furthermore, the consistency of our $\Delta_i$ values with those obtained from magnetic measurements validates our findings.

\subsection{Resistivity}

\begin{figure}[ht!]
    \centering
    \includegraphics[width=0.9\columnwidth]{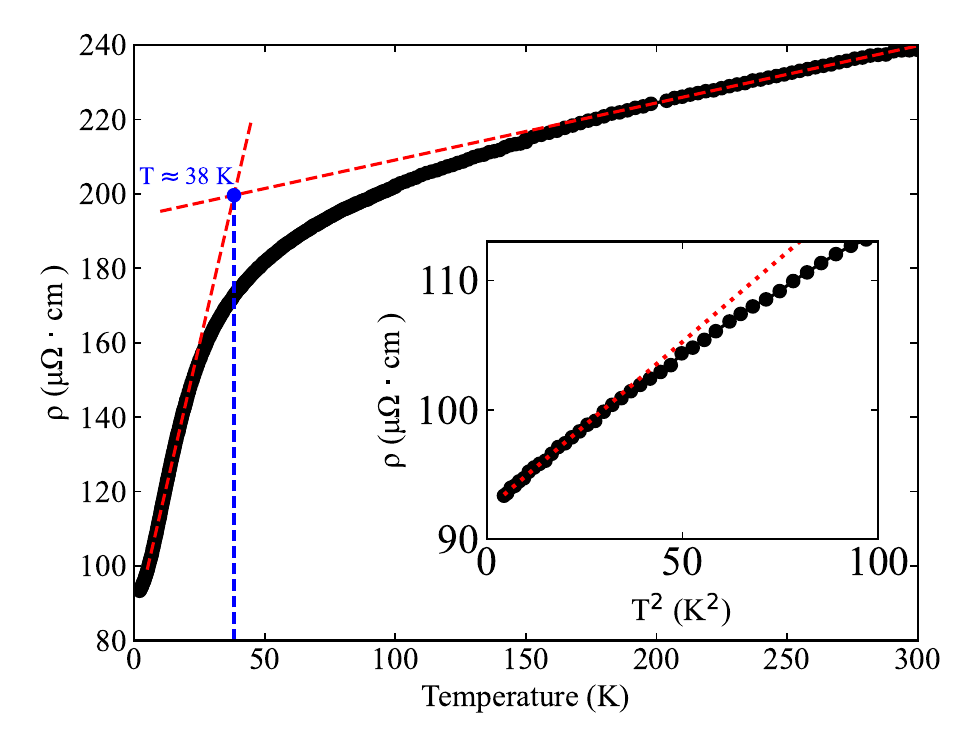}
    \caption{Temperature dependence of the resistivity for YbV$_6$Sn$_6$. The red dashed lines represent linear fits before and after the Kondo coherence temperature, whose intersection defines the Kondo temperature (\textit{$T_{\text{K}}$}). The blue dashed line indicates the determined \textit{$T_{\text{K}}$}. The inset shows the T$^2$ dependence of the resistivity in low temperatures and the Fermi-liquid model fitting from 2 to 6~K.}
    \label{fig:GMQ_926_0GPa_0kOe}    
\end{figure}

The resistivity profile shown in Fig.~\ref{fig:GMQ_926_0GPa_0kOe} exhibits the characteristic behavior of a Kondo lattice, as evidenced by a shoulder-like decrease below the Kondo coherence temperature.
The obtained Kondo temperature \textit{$T_{\text{K}} = 38.25$ K} is comparable with the value of 30~K reported in the literature \cite{YbV6Sn6}.

In the low-temperature regime (2~K to 6~K), the Fermi-liquid model fitting yielded a resistivity coefficient $A = 0.200(2)$~$\mu\Omega\cdot$cm$\cdot$K$^{-2}$, leading to a Kadowaki-Woods ratio $A/\gamma = 2.15(2) \times 10^{-6}$~$\mu\Omega\cdot$cm$\cdot$K$^{2}\cdot$mol$^{2}\cdot$mJ$^{-2}$.
This ratio approaches the expected value for a Yb$^{3+}$ ion degeneracy ($N=4$) \cite{kadowaki1986universal, Generalized-KW}.
These results are consistent with the CEF estimations, since the fitting temperature range corresponds exclusively to the first excited state energy level ($\Delta_1 = 6.702$ K).

\subsubsection{Magnetic Field Dependence}
\begin{figure}[ht!]
    \centering
    \includegraphics[width=0.9\columnwidth]{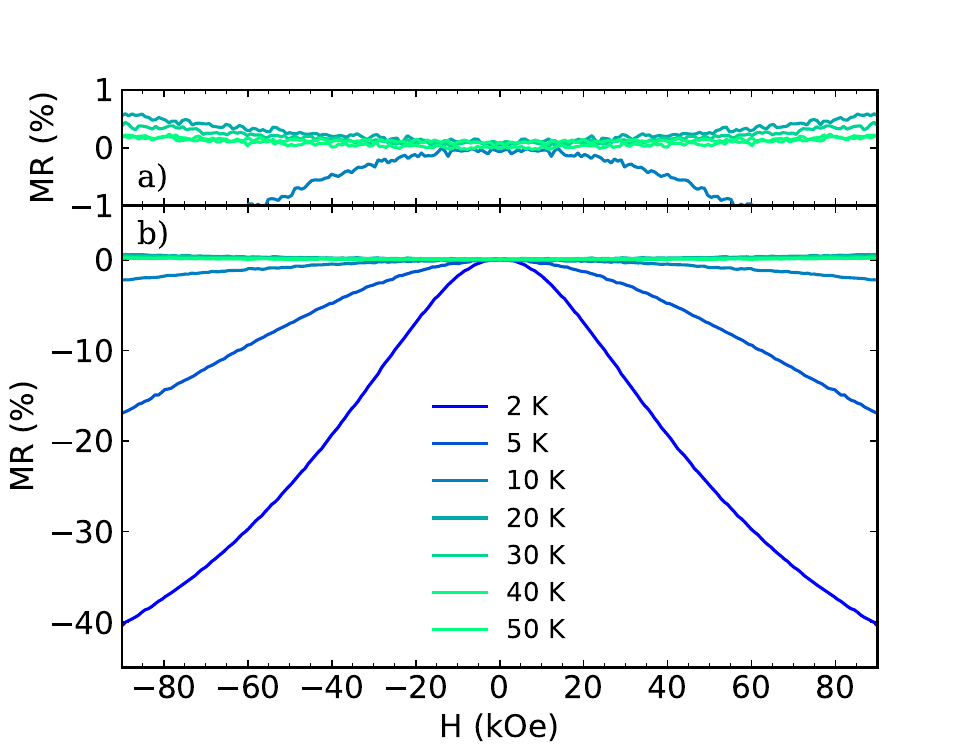}
    \caption{Relative magnetoresistance of YbV$_6$Sn$_6$ as a function of magnetic field at various temperatures. (a) MR curves for temperatures from 20 K to 50 K, where the magnetoresistance is positive for all fields. Notably, a change in concavity is observed at 10 K. (b) MR curves for temperatures from 2 K to 10 K, where the magnetoresistance is negative for all fields, exceeding 40\% at 2 K and 90 kOe.}
    \label{fig:GMQ_926_MR}    
\end{figure}

The magnetoresistance of YbV$_6$Sn$_6$ was measured at fixed temperatures 2, 5, 10, 20, 30, 40 and 50~K while varying the magnetic field from -90~kOe to 90~kOe with $i \parallel ab \perp H$.
Figure~\ref{fig:GMQ_926_MR} shows the relative magnetoresistance, defined as $MR = [R(H)-R(0)]/R(0)$.
The magnetoresistance below 10~K is negative for all fields, exceeding 40\% at 2~K and for 90~kOe.
For $T \geq$ 20~K, the magnetoresistance is positive for all fields (see panel of Fig.~\ref{fig:GMQ_926_MR}).
Similar behavior was found in systems such as Yb$_4$TGe$_8$ (T = Cr, Mn, Fe, Co, and Ni) \cite{PhysRevB.106.024402}.
In such systems, the destabilization of the Kondo lattice can cause positive magnetoresistance, while the suppression of additional magnetic correlations can cause negative magnetoresistance.
It is important to note that recent $^{51}$V NMR measurements in YbV$_6$Sn$_6$ indicated the suppression of spin fluctuations around 20~K for $H \parallel c$ \cite{park2025investigationparamagneticstatekagome}, the same region where the magnetoresistance shows a change in  concavity.

\subsubsection{Pressure Dependence}

\begin{figure}[ht!]
    \centering
    \includegraphics[width=0.9\columnwidth]{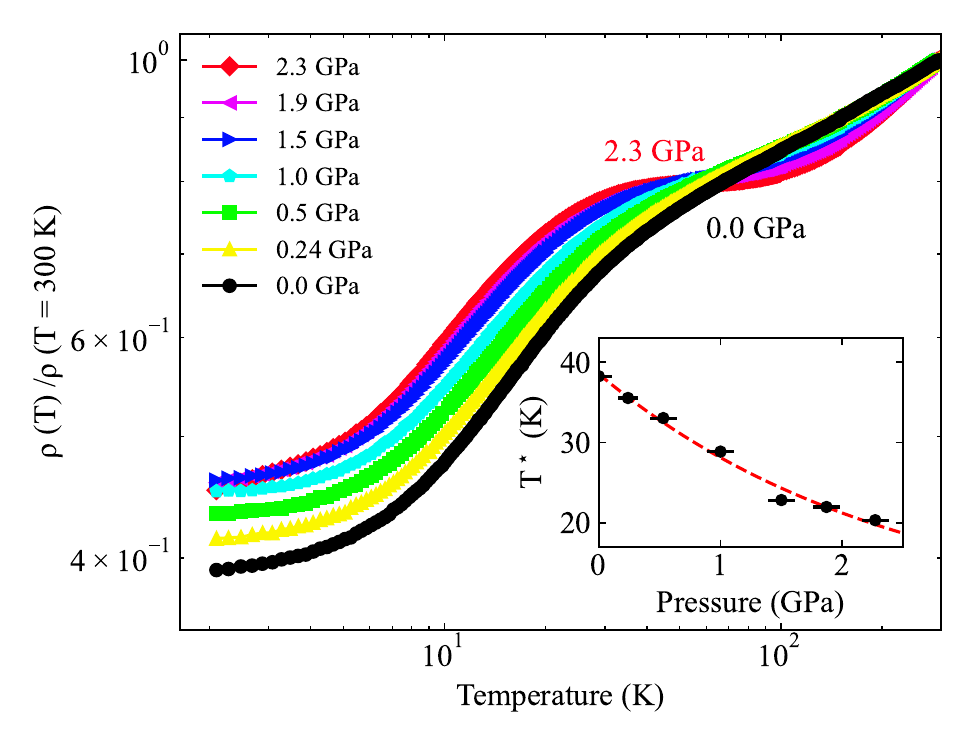}
    \caption{Normalized electrical resistivity, $\rho$(T)/$\rho$(T=300~K), as a function of temperature under various hydrostatic pressures. Data are shown from 0.0 GPa (black circles) up to 2.3 GPa (red circles), illustrating the shift in the onset of the shoulder-like resistivity drop. Inset: Pressure dependence of the characteristic temperature \textit{T}$^{\star}$, defined as the initiation of the resistivity drop, revealing a continuous decrease with increasing pressure. The dashed red line is a guide to the eye.}
    \label{fig:GMQ_926_Tk}    
\end{figure}

Figure \ref{fig:GMQ_926_Tk} shows the temperature dependence of the resistivity measured under applied pressures ($0 \leq P \leq 2.3$~GPa) in the temperature range from 2~K to 300~K with $i \parallel ab$.
Normalizing the resistivity curves by their respective magnitudes at 300~K clearly demonstrates a shift in the onset of the shoulder-like resistivity drop towards lower temperatures.
This observation is consistent with the Doniach diagram for Yb-based compounds, where pressure application is expected to reduce the Kondo temperature ($T_K$) \cite{doniach1977kondo}.
Above 1.0~GPa, a negative concavity emerges in the resistivity around 80~K just before the drop, revealing the characteristic broad peak from the Kondo coherence temperature.
The prominence of this negative concavity, followed by a local maximum, likely arises from a reduction in the phononic contribution to the resistivity under pressure.
This reduction unmasks the intrinsic resistivity response dominated by Kondo scattering.
Given the difficulty in unambiguously identifying the local resistivity maximum as the coherent Kondo temperature in this system, particularly at low pressures, we defined a reference temperature, $T^{\star}$, marking the beginning of the resistivity drop.
The pressure dependence of $T^{\star}$ is plotted in the inset of Fig.~\ref{fig:GMQ_926_Tk}, revealing behavior analogous to the expected pressure dependence of $T_K$.

As pressure increases up to 1.0 GPa, a subtle change in the low-temperature (2 to 10 K) resistivity concavity is accompanied by an increase in residual resistivity, $\rho_0$. Beyond that pressure, the response tends towards linearization, concurrent with a decrease in $\rho_0$.
Such behavior is intimately linked to a modification in the scattering mechanisms of the conduction electrons and warrants analysis using a power law, similar to the approach taken under an external magnetic field.

Analysis of the exponent $n$ reveals that the system initially starts as a non-Fermi-liquid (NFL) state, which suggests proximity to the previously reported quantum critical point, easily tuned by magnetic field \cite{YbV6Sn6}.
Furthermore, interpreting this result within the framework of the Doniach diagram indicates that the analyzed region lies on the right side of the antiferromagnetic (AFM) phase.
Consequently, we anticipate that an increase in pressure will lead to an initial increase in $T_N$, reaching a maximum, and then slowly decreasing towards $T_N=0$.
This decrease would stem from the vanishing of exchange interaction between conduction electrons and the localized Yb $4f$ electrons.

\begin{figure}[ht!]
    \centering
    \includegraphics[width=0.9\textwidth]{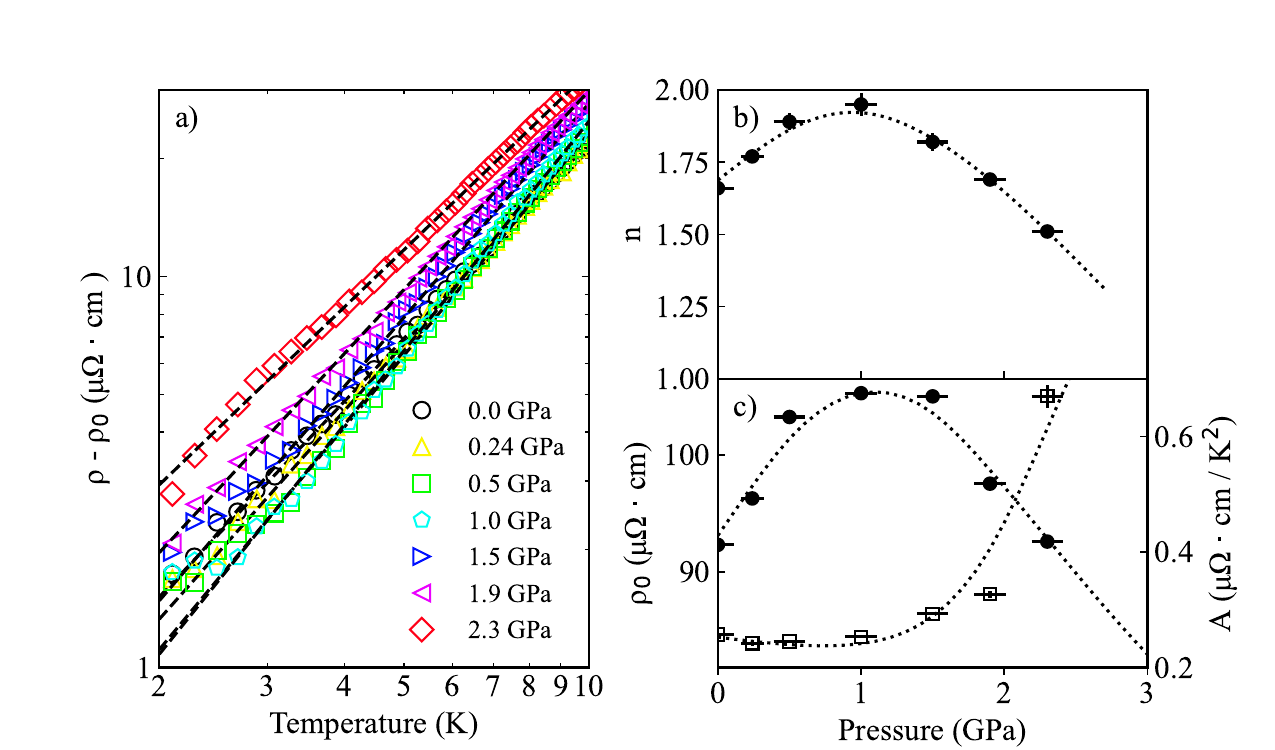}
    \caption{(a) Temperature dependence of the resistivity $\rho - \rho_0$  on a logarithmic scale for various pressures, along with power law fits $\rho - \rho_0 = AT^n$. The symbols represent experimental data for different pressures from 0.0 GPa (black circles) to 2.3 GPa (red diamonds), and the dashed lines are the corresponding fits. (b) Pressure dependence of the exponent $n$, extracted from the power law fits in (a). (c) Pressure dependence of the residual resistivity $\rho_0$ (black circles, left axis) and the coefficient $A$ (black squares, right axis), also obtained from the fits in (a). Dotted lines in (b) and (c) are guides to the eye.}
    \label{fig:GMQ_926_Ajuste_low_pressao}    
\end{figure}

The exponent $n$ then gradually increases towards a Fermi-liquid (FL) state ($n=2$) until approximately 1.0~GPa.
However, our measurements down to 2~K did not provide clear indications of a magnetic transition, thus requiring measurements at even lower temperatures to definitively confirm the increase of $T_N$ and a potential $T_N^{max}$ around 1.0~GPa.
For pressures exceeding 1.0 GPa, the exponent $n$ decreases systematically (indicated by the dashed line in Fig.~\ref{fig:GMQ_926_Ajuste_low_pressao}~b).
This signature indicates possible entry into a second quantum critical regime after exiting the previous quantum critical region by reaching FL character around 1~GPa.
A linear extrapolation of the trend between 1~GPa and 2.3~GPa predicts that $n = 1$ is reached around 3.0~GPa.
Consistently, the residual resistivity $\rho_0$ mirrors the behavior described for $n$, diverging from the general pressure dependence of the total resistivity, which tends to decrease with increasing pressure.
This discrepancy strongly suggests the presence of additional mechanisms influencing electron conduction at low temperatures.

In particular, the $A$ parameter exhibits an inverse behavior to that of $n$.
It is noteworthy that a decrease in $\rho_0$ accompanied by an increase in $A$ has been previously observed in YbIr$_2$Si$_2$ \cite{yuan2006quantum} and YbNiGe$_3$ \cite{umeo2010pressure, sato2014pressure, sato2015electronic}, being associated with changes in the valence of Yb.

\section{Conclusions}

In conclusion, our experimental study of the low-temperature electrical resistivity of YbV$_6$Sn$_6$ reveals a significant complexity in its phase diagram that goes beyond the classical Doniach model.

From our measurements under pressure, analysis of the exponent $n$ initially points to an NFL state.
The application of pressures drives the system towards an FL state ($n=2$) up to approximately 1.0~GPa, a distinct systematic decrease in $n$ for pressures exceeding 1.0~GPa suggests the emergence of a second QCP, with $n=1$ estimated around 3.0~GPa.
The anomalous behavior of the residual resistivity ($\rho_0$), which aligns with that of $n$ rather than the general trend of decreasing total resistivity with pressure, further corroborates the existence of complex scattering mechanisms at low temperatures.

These findings suggest a complex putative phase diagram for YbV$_6$Sn$_6$, where distinct quantum critical points are tuned by magnetic field and pressure.
However, for a complete understanding of this phenomenon, it's crucial to consider the complex nature of the material's interactions, where the Kondo lattice effect, RKKY interaction, and CEF effects operate on very close energy scales.
Further measurements - including specific heat, magnetic susceptibility, and inelastic neutron scattering under pressure -  could provide valuable insights into the underlying mechanisms.
This would allow a direct evaluation of how the CEF acts in the sensitive competition between magnetic ordering and Kondo screening as a function of pressure.
This material, therefore, serves as a promising platform for exploring multi-parameter quantum criticality and the interplay between Kondo physics, magnetism and CEF.

\begin{acknowledgments}
We acknowledge the financial support of Brazilian funding agencies CAPES, CNPq (Contracts No. 140921/2022-2, No. 88887.837417/2023-00), FAPESP (Contracts No. 2017/20989-8, No. 2017/10581-1, No. 311783/2021-0, No. 405408/2023-4), and the experimental support from the Multiuser Central Facilities (UFABC) and LCCEM (UFABC).
\end{acknowledgments}


\bibliography{apssamp}

\end{document}